\documentclass{elsart1p}
\usepackage{graphicx}
\usepackage{amssymb}
\def\NIMA#1#2#3{Nucl. Inst. Methods {\bf A#1} (#2) #3}
\begin{document}
\begin{frontmatter}
%
%
%
\title{Measurements of CP violation parameters\\at the NA48 experiment at CERN}
%
%
\author{Evgueni Goudzovski}
\address{School of Physics and Astronomy, University of Birmingham, B15 2TT, United Kingdom}
\begin{abstract}
Recent precise measurements of CP violation parameters in kaon
decays at the NA48 experiment: indirect CPV parameter $|\eta_{+-}|$,
and charge asymmetries in $K^\pm\to3\pi$ decays, are presented.
\end{abstract}
\end{frontmatter}
%
\section{Introduction}
\label{intro}
The CERN programme in experimental kaon physics of the last decade
has been carried out by the NA48 series of experiments. NA48 has
accomplished several physics subprogrammes based on data taken with
$K_L$, $K_S$, $K^\pm$ and neutral hyperon beams in 1997--2004. The
principal components of the experimental setup (modified and
upgraded in the course of operation) are a beam line followed by a
vacuum decay volume, a magnetic spectrometer consisting of four
drift chambers, a trigger scintillator hodoscope, a liquid krypton
electromagnetic calorimeter, a hadron calorimeter, and and a muon
detector~\cite{fa07}.

The present paper reports a number recent precise measurements of CP
violation (CPV) parameters: 1) the indirect CPV parameter
$|\eta_{+-}|$ with $K_L\to\pi^+\pi^-$ decays; 2) the direct CP
violating charge asymmetries of Dalitz plot slopes $A_g$ in
$K^\pm\to3\pi^\pm$ and $K^\pm\to\pi^\pm\pi^0\pi^0$ decays.
\boldmath
\section{Measurement of the indirect CP violation parameter
$|\eta_{+-}|$} \unboldmath \label{eta}
The interest in a precise measurement of the indirect CPV parameter
$|\eta_{+-}|=A(K_L\to\pi^+\pi^-)/A(K_S\to\pi^+\pi^-)$ stems, in
particular, from the fact that its recent measurements by KTeV and
KLOE experiments published in 2004 and 2006, respectively, differ by
$\sim\!5\%$, or more than four standard deviations, from the
previous world average.

The NA48 measurement of $|\eta_{+-}|$~\cite{la07} is based on a data
set taken during two days of dedicated running in 1999. The directly
measured quantity is the ratio of the decay rates
$R=\Gamma(K_L\to\pi^+\pi^-)/\Gamma(K_L\to\pi e\nu)$; these decays
are characterized by similar signatures involving two reconstructed
tracks of charged particles. Then $|\eta_{+-}|$ is computed as
\begin{equation}
\label{eta-main} |\eta_{+-}| =
\sqrt{\frac{\Gamma(K_L\to\pi^+\pi^-)}{\Gamma(K_S\to\pi^+\pi^-)}}=
\sqrt{\frac{\textrm{BR}(K_L\to\pi^+\pi^-)}{\textrm{BR}(K_S\to\pi^+\pi^-)}
\cdot\frac{\tau_{KS}}{\tau_{KL}}}.
\end{equation}
In this approach the $K_L$ and $K_S$ lifetimes $\tau_{KL}$ and
$\tau_{KS}$, and the branching fractions $\textrm{BR}(K_L\to\pi
e\nu)$ and $\textrm{BR}(K_S\to\pi^+\pi^-)$ are external inputs taken
from the best single measurements.

The data sample contains about $80\times10^6$ 2-track triggers.
Event selection is similar for the $K_L\to\pi^+\pi^-$ and $K_L\to\pi
e\nu$ modes. A crucial difference is electron vs pion identification
based on the ratio of particle energy deposition in the EM
calorimeter to its momentum measured by the spectrometer (expected
to be close to 1 for electrons). Particle identification
efficiencies were directly measured and corrected for.

Samples of $47\times 10^3$ $K_L\to\pi^+\pi^-$ and $5.0\times 10^6$
$K_L\to\pi e\nu$ candidates were selected, with about $0.5\%$
background contamination in each. Acceptance corrections and
background subtraction were performed by Monte Carlo simulation.
Trigger efficiencies were measured directly with the data and
corrected for. The most relevant systematic uncertainties come from
precision of simulation of kaon momentum spectrum, precision of
radiative corrections, and precision of trigger efficiency
measurement. The final result is
\begin{equation}
\Gamma(K_L\to\pi^+\pi^-)/\Gamma(K_L\to\pi e\nu) =
(4.835\pm0.022_{stat.}\pm0.016_{syst.})\times 10^{-3}.
\end{equation}
This leads, subtracting the $K_L\to\pi^+\pi^-\gamma$ direct emission
contribution, but retaining the inner bremsstrahlung contribution,
to
\begin{equation}
\textrm{BR}(K_L\to\pi^+\pi^-) = (1.941\pm0.019)\times 10^{-3}.
\end{equation}
Finally, the CP violation parameter is computed according to
(\ref{eta-main}) to be
\begin{equation}
|\eta_{+-}| = (2.223\pm0.012)\times 10^{-3}.
\end{equation}
The result in in agreement with the recent KLOE and KTeV
measurements, while it contradicts the 2004 PDG average. The latter
disagreement is understood to be due to the improved treatment of
the radiative corrections in the recent analyses.
\boldmath
\section{Measurement of the direct CPV parameter $A_g$ in
$K^\pm\to3\pi$ decays} \unboldmath

$K^\pm\to\pi^\pm\pi^+\pi^-$ and $K^\pm\to\pi^\pm\pi^0\pi^0$ decays
are among the most promising processes in kaon physics to search for
CPV phenomena. The decay density is parameterized (up to radiative
and $\pi\pi$ rescattering effects studied
separately~\cite{slopes,cusp}) by a polynomial expansion
\begin{equation}
d^2\Gamma/dudv\sim 1+gu+hu^2+kv^2, \label{slopes}
\end{equation}
where $g$, $h$, $k$ are the so called linear and quadratic Dalitz
plot slope parameters ($|h|,|k|\ll |g|$), and the two Lorentz
invariant kinematic variables $u$ and $v$ are defined as
\begin{equation}
u=\frac{s_3-s_0}{m_\pi^2},~~v=\frac{s_2-s_1}{m_\pi^2},~~
s_i=(P_K-P_i)^2,~i=1,2,3;~~s_0=\frac{s_1+s_2+s_3}{3}. \label{uvdef}
\end{equation}
Here $m_\pi$ is the charged pion mass, $P_K$ and $P_i$ are the kaon
and pion four-momenta, the indices $i=1,2$ correspond to the two
pions of the same electrical charge, and the index $i=3$ to the pion
of different charge. A non-zero difference $\Delta g$ between the
slope parameters $g^+$ and $g^-$ describing the decays of $K^+$ and
$K^-$, respectively, is a manifestation of direct CP violation
expressed by the corresponding slope asymmetry
\begin{equation}
A_g = (g^+ - g^-)/(g^+ + g^-) \approx \Delta g/(2g). \label{agdef}
\end{equation}
The above slope asymmetry is expected to be strongly enhanced with
respect to the asymmetry of integrated decay rates. A recent full
next-to-leading order ChPT computation~\cite{ga03} predicts $A_g$ to
be of the order of $10^{-5}$ within the SM. Calculations involving
processes beyond the SM~\cite{sh98,ga00} allow a wider range of
$A_g$, including substantial enhancements up to a few $10^{-4}$.

A measurement of the quantity $A_g$ was performed with a record data
sample collected in 2003--04 with simultaneous $K^+$ and $K^-$
beams~\cite{ba07}. The measurement method is based on the study of
ratios of $u$ spectra of $K^+$ and $K^-$ decays, and exploits
cancellations of major systematic effects due to the simultaneous
collection of $K^+$ and $K^-$ decays, and regular inversions of
magnetic fields in the beam line and the spectrometric magnet, which
allows achieving $\sim10^{-4}$ precision. The event samples are
practically background-free, and contain $3.11\times 10^9$
$K^\pm\to\pi^\pm\pi^+\pi^-$ candidates, and $9.13\times 10^7$
$K^\pm\to\pi^\pm\pi^0\pi^0$ candidates (the $K^+$/$K^-$ flux ratio,
on which however the results do not depend, is 1.8).

The CP violating charge asymmetries of the linear slope parameter of
the Dalitz plot of the $K^\pm\to\pi^\pm\pi^+\pi^-$ and
$K^\pm\to\pi^\pm\pi^0\pi^0$ decays were measured to be
\begin{equation}
\begin{array}{rcrllllcrcl}
A_g^c &=& (-1.5 &\pm& 1.5_{stat.} &\pm& 1.6_{syst.})\times
10^{-4} &=& (-1.5&\pm&2.2)\times 10^{-4},\\
A_g^n &=& (1.8 &\pm& 1.7_{stat.}&\pm& 0.6_{syst.})\times 10^{-4}&=&
(1.8&\pm&1.8)\times 10^{-4}.
\end{array}
\end{equation}
The archived precision is more than an order of magnitude better
that those of the previous measurements. The results do not show
evidences for large enhancements due to non-SM physics, and can be
used to constrain certain SM extensions predicting enhanced CPV
effects.

\section{Summary}
A number of recent measurements of CPV parameters in kaon decays by
the NA48 collaboration at CERN are presented. The achieved
precisions are similar to or better than the best previous ones.

\end{document}